\documentclass[nofootinbib,letterpaper,aps,amsmath,amsfonts,twocolumn]{revtex4}
\begin{document}
\newcommand{\beeq}{\begin{equation}}
\newcommand{\eneq}{\end{equation}}
\newcommand{\beeqar}{\begin{eqnarray}}
\newcommand{\eneqar}{\end{eqnarray}}
\title{A new form of the no-cloning theorem for harmonic oscillator coherent states. 
}
\author{N.D. Hari Dass\ddag} 
\address{Centre for High Energy Physics, Indian Institute of Science, Bangalore 560012, 
India}
\begin{abstract}
We give an alternative formulation of the no-cloning theorem that applies to harmonic oscillator coherent states.
It says that {\em unknown} single harmonic oscillator coherent states  can not be {\em amplified}. Conversely it says that
{\em known} harmonic oscillator coherent states can be directly amplified through an universal unitary process.
It is also shown that such a formulation is needed for the consistency of {\em information cloning} proposed by us
long ago.
\end{abstract}
\maketitle
\vskip 2pc
\section{Introduction}
{\label{intro}}

The well known {\it no cloning theorem } \cite{noclone} which asserts
that in quantum theory it is {\em impossible} to
make copies of an {\em unknown} single state is a remarkable aspect of
quantum theory. 
It provides a remarkable consistency to the
ensemble interpretation. For, if such a copying
were possible, one could have created an 
{\it arbitrarily large} ensemble and determined the
statistical significance through {\em ensemble
measurements}. What is remarkable about this consistency is the
fact that while one deals with unitary evolution, the other deals with
quantum measurements!

In this brief note we give a somewhat more physically comprehendable
formulation of the no-cloning theorem that is applicable to harmonic oscillator
coherent states.

In this formulation the no-cloning theorem can be stated as 'It is impossible
to {\em amplify} the excitation level of a single unknown harmonic oscillator
coherent state. The single nature is crucial as if we had an ensemble of such
states we could determine the state and apply suitable unitary transformations
to obtain a state with scaled excitation. Of course there is no guarantee that such a
unitary process would be {\em universal}.

In fact the following argument might indicate that such a universal process can not
exist. Let us suppose we have two coherent states $|\alpha\rangle, |\beta\rangle$
and let an universal unitary process $U$ change their excitation (or equivalently
the coherence parameters $\alpha,\beta$ by a common (complex) scale $\lambda$.
Then
\beeqar
U|\alpha\rangle &=& |\lambda\cdot\alpha\rangle~\nonumber\\
U|\beta\rangle  &=& |\lambda\cdot\beta\rangle
\eneqar
It then follows that
\beeq
|\langle\alpha|\beta\rangle|^2 = |\langle\lambda\cdot\alpha|\lambda\cdot\beta\rangle|^2
\eneq
On noting the overlap for coherent states $|\langle\alpha|\beta\rangle|^2 = e^{-|\alpha-\beta|^2}$
it follows that $\lambda = 1$. This argument, though correct in itself, is not sufficient for our
purposes. Not unexpectedly ancillary states are required to show that excitations can indeed be scaled.
The {\em information cloning} protocol given by us long ago \cite{incl} is precisely what is required.
That protocol can be stated as follows:

Consider a {\em single unknown} coherent state $|\alpha\rangle$ and an ensemble of {\em known} coherent
states which can also be displayed as a disentangled product state
\beeq
|\Psi\rangle = |\alpha\rangle\cdot|\beta\rangle_1\cdot|\beta\rangle_2\cdots|\beta\rangle_N
\eneq

The $1+N$ harmonic oscillators are described by the set of
creation and annihilation operators $(a,a^\dag),
(b_k,b_k^\dag)$ (where the index $k$ takes on values $1,..,N$)
satisfying the commutation relations
\beeq \label{4}
[a,a^\dag]~=~1;~~[b_j,b_k^\dag]~=~\delta_{jk};~~[a,b_k]~=~0;~~[a^\dag,b_k]~=~0
\eneq
The (information)-cloning transformation is the universal unitary operator
\beeq \label{7}
U = e~~^ {~-\frac{\pi}{2\sqrt{N}}(a^\dag\otimes\sum_j r_j b_j - a\otimes\sum_j r_j b_j^\dag)}
\eneq
The result of the operation of $U$ on $|\Psi\rangle$ is given by
\beeq
\label{inclone}
U|\Psi\rangle = |-\sqrt{N}\beta\rangle\cdot|\frac{\alpha}{\sqrt{N}}\rangle_1\cdots|\frac{\alpha}{\sqrt{N}}\rangle_N
\eneq
This is what we have called {\em information cloning} earlier \cite{incl}.
For coherent states complete
information is contained in the parameter $\alpha$. The original unknown single state $|\alpha\rangle$ has now been
cloned into $N$ identical copies of the state $|\frac{\alpha}{\sqrt{N}}\rangle$. By information cloning what we mean is this
ability to make arbitrary number of copies of coherent states 
whose coherency parameter is $c(N)\alpha$ where $\alpha$ is the coherency
parameter of the unknown coherent state and $c(N)$ is a known constant
depending on the number of copies made. It is not cloning because the $N$-copies are not identical to the
original unknown state; yet the copies carry all the information about the original unknown state.

To elucidate matters further let us consider the action of $U$ on another state $|\Psi^\prime\rangle$:
\beeq
|\Psi^\prime\rangle = |\alpha^\prime\rangle\cdot|\beta^\prime\rangle_1\cdot|\beta^\prime\rangle_2\cdots|\beta^\prime\rangle_N
\eneq
Then
\beeq
U|\Psi^\prime\rangle = |-\sqrt{N}\beta^\prime\rangle\cdot|\frac{\alpha^\prime}{\sqrt{N}}\rangle_1\cdots|\frac{\alpha^\prime}{\sqrt{N}}\rangle_N
\eneq

We can then compare the squared inner products $|\langle \Psi|\Psi^\prime\rangle|^2 $ and $|\langle U\Psi|U\Psi^\prime\rangle|^2$:
\beeqar
|\langle\Psi|\Psi^\prime\rangle|^2 &=& |\langle\alpha|\alpha^\prime\rangle|^2\cdot|\langle\beta|\beta^\prime\rangle|^{2N}\nonumber\\
                                   &=& e^{-|\alpha-\alpha^\prime|^2}\cdot e^{-N|\beta-\beta^\prime|^2}\nonumber\\
|\langle U\Psi| U\Psi^\prime\rangle|^2 &=& |\langle -N\beta|-N\beta^\prime\rangle|^2\cdot|\langle\frac{\alpha}{\sqrt{N}}|{\frac{\alpha}{\sqrt{N}}}^\prime\rangle|^{2N}\nonumber\\
                                   &=& e^{-N|\beta-\beta^\prime|^2}\cdot e^{-N\cdot\frac{1}{N}|\alpha-\alpha^\prime|^2}
\eneqar
Obviously the two expressions agree. It is worth examining closely the manner in which they agree with each other. The excitation of the 
unknown single state $|\alpha\rangle$ has been split evenly among the $N$-states $|\frac{\alpha}{\sqrt{N}}\rangle$; the excitations of the
$N$ known states $|\beta\rangle$ have been merged into the excitation of the single known state $|-N\beta\rangle$. Thus the information
cloning transformation of eqn(\ref{7}) has {\em attenuated} the single unknown state while at the same time amplified the known state.

The latter fact gives rise to the converse statement of the new form of the no-cloning theorem for harmonic oscillator coherent states:'It
is possible to amplify known harmonic oscillator coherent states'. We see that in fact N-copies of such a known state are required. Since the
state is known there is no problem in producing N identical copies of it.

If the new formulation of the no-cloning theorem was not correct, we would immediately run into a contradiction. Suppose, in opposition to
the new form of the no-cloning theorem, it were possible to amplify an unknown coherent state through an universal unitary process(one is
allowed to use further ancillaries), then the states $|\frac{\alpha}{\sqrt{N}}\rangle$ could have been amplified to $|\alpha\rangle$. But then
we would have succeeded in cloning the unknown single state $|\alpha\rangle$! This would have violated the no-cloning theorem. Of course the transformation of eqn(\ref{7}) did manage to amplify the state $|\beta\rangle$, but that was a {\em known} state.

We finally return to the consistency that the original no-cloning theorem gave to the ensemble interpretaion of quantum mechanics.
Will the information cloning upset this consistency? This issue has been analyzed in \cite{singquant}. It was found there that information
cloning enables to determine the coherency parameter of the unknown coherent state via an ensemble measurement on the N-copies 
$|\frac{\alpha}{\sqrt{N}}\rangle$ where the average value of the measurement correctly estimates $\alpha$
\beeq \label{46}
\alpha_{est} = \alpha
\eneq
but the measured variances are given by
\beeq
\Delta \alpha_R = \Delta \alpha_I = \frac{1}{\sqrt{2}}
\eneq
Thus, while the statistical error in usual measurements goes as $\frac{1}{\sqrt{N}}$,
and can be made arbitrarily small by making $N$ large enough, information cloning gives
an error that is fixed and equal to the quantum mechanical uncertainty associated with the original unknown state.

\section{Acknowledgements}
The author would like to express his gratitude to the Department of Atomic Energy for the award of a
Raja Ramanna Fellowship which made this work possible, and to CHEP, IISc for its invitation to use
this Fellowship there.

\end{document}